\documentclass[conference]{IEEEtran}
\IEEEoverridecommandlockouts

\usepackage{balance}
\usepackage{cite}
\usepackage{amsmath,amssymb,amsfonts}
\usepackage{algorithmic}
\usepackage{graphicx}
\usepackage{textcomp}
\usepackage{xcolor}
\def\BibTeX{{\rm B\kern-.05em{\sc i\kern-.025em b}\kern-.08em
    T\kern-.1667em\lower.7ex\hbox{E}\kern-.125emX}}
\begin{document}

\title{Single Channel Speech Enhancement Using \\
U-Net Spiking Neural Networks}

\author{\IEEEauthorblockN{Abir Riahi}
\IEEEauthorblockA{\textit{Department of Electrical and Computer Engineering} \\
\textit{Université de Sherbrooke}\\
Sherbrooke, Quebec, Canada \\
abir.riahi@usherbrooke.ca}
\and
\IEEEauthorblockN{Éric Plourde}
\IEEEauthorblockA{\textit{Department of Electrical and Computer Engineering} \\
\textit{Université de Sherbrooke}\\
Sherbrooke, Quebec, Canada \\
eric.plourde@usherbrooke.ca}
}

\maketitle

\begin{abstract}
Speech enhancement (SE) is crucial for reliable communication devices or robust speech recognition systems. Although conventional artificial neural networks (ANN) have demonstrated remarkable performance in SE, they require significant computational power, along with high energy costs. In this paper, we propose a novel approach to SE using a spiking neural network (SNN) based on a U-Net architecture. SNNs are suitable for processing data with a temporal dimension, such as speech, and are known for their energy-efficient implementation on neuromorphic hardware. As such, SNNs are thus interesting candidates for real-time applications on devices with limited resources. The primary objective of the current work is to develop an SNN-based model with comparable performance to a state-of-the-art ANN model for SE. We train a deep SNN using surrogate-gradient-based optimization and evaluate its performance using perceptual objective tests under different signal-to-noise ratios and real-world noise conditions. Our results demonstrate that the proposed energy-efficient SNN model outperforms the Intel Neuromorphic Deep Noise Suppression Challenge (Intel N-DNS Challenge) baseline solution and achieves acceptable performance compared to an equivalent ANN model.
\end{abstract}

\begin{IEEEkeywords}
speech enhancement, spiking neural network, surrogate gradient
\end{IEEEkeywords}

\section{Introduction}\label{sec:introduction}

Speech enhancement (SE) is an important signal processing task that improves speech quality by removing background additive noises while maintaining speech intelligibility. SE is essential for many applications such as speech recognition, communication devices, video conferencing systems and hearing aids.

Classic SE techniques such as spectral subtraction \cite{Boll}, wiener filtering \cite{McAulay}, minimum mean square error estimation \cite{Ephraim}, and subspace methods \cite{Asano} have been widely used and have shown to be effective in certain situations. However, they have limitations when dealing with non-stationary noise and low Signal-to-Noise Ratio (SNR) conditions. Deep learning models, on the other hand, have shown promising results in addressing these challenges by leveraging large amounts of training data to learn complex mappings between noisy and clean speech signals \cite{Xu}.

In fact, in recent years, deep learning architectures demonstrated great potential to deploy robust SE systems in real-world environments \cite{Hu}. The success of conventional artificial neural networks (ANN) can be attributed to the availability of large datasets and high-performance computational resources. However, most SE applications require real-time computation on limited resource devices such as cellphones, which poses a significant challenge to deploy ANNs. In fact, these networks generally require significant computational resources and massive energy consumption.
   
Spiking neural networks (SNN), which are low-power deep neural networks, have emerged as a promising alternative to ANNs. SNNs offer comparable computational complexity to ANNs while consuming significantly less energy. The asynchronous, binary, and sparse event-driven processing of SNNs is one of the primary reasons behind their energy efficiency on neuromorphic hardware. SNN training poses a significant challenge due to the non-differentiable nature of its spiking function. Consequently, extensive research efforts have been directed towards exploring novel SNN learning methods. Recently, a novel technique known as surrogate gradient optimization \cite{Neftci} has demonstrated remarkable robustness in training deep SNN architectures using backpropagation on a wide range of tasks, including classification \cite{Zenke}, speech command recognition \cite{Pellegrini}, depth estimation \cite{Rançon}, among others.

However, a very limited number of research works have been done for SE using SNNs \cite{Wall,Xing,Timchek}. Among these, Wall \textit{et al.} \cite{Wall} propose a three-layer SNN architecture that employs lateral inhibition. To generate the training dataset, the authors introduce three levels of additive Gaussian white noise and subsequently compute the Short Time Fourier Transform (STFT) of the signal. The log scaled STFT magnitude is then encoded into discrete spike timing using the Bens Spiker Algorithm (BSA) \cite{Schrauwen}. The proposed SNN model then processes the encoded input spikes using a masking approach to eliminate uncorrelated spikes. In other words, the enhanced STFT is obtained by performing an element-wise multiplication of the complex noisy STFT with the SNN's output spike train. The results of their experiments demonstrate favorable performance in terms of SNR. However, the proposed approach has some limitations. Specifically, the SNN utilized in this study does not incorporate any form of learning, and the architecture is relatively shallow. Furthermore, the proposed model could have been subjected to more extensive testing using a diverse array of noise types to validate its generalizability. Xing \textit{et al.} \cite{Xing} propose a similar approach by leveraging a three-layer SNN architecture that incorporates lateral inhibition. The authors utilize a log scaled STFT magnitude as the input current for the SNN model, which then generates a binary mask. The enhanced STFT is obtained by computing an element-wise multiplication with the binary mask. Their experimental design includes the incorporation of five distinct real-world noise types, and the resulting SE model demonstrates good performance with respect to SNR. However, the SE process relies on the elimination of uncorrelated noise components, and the SNN architecture does not integrate any learning strategies. A more recent work by Intel developed a simple SNN-based baseline solution in the context of the Intel Neuromorphic Deep Noise Suppression Challenge (Intel N-DNS Challenge) \cite{Timchek}. The proposed method utilizes a three-layer feedforward sigma-delta neural network (SDNN) to mask the STFT Magnitude. The delta-encoded STFT magnitude is employed as the input to the SNN, which generates a multiplicative mask for computing the enhanced STFT. The baseline SNN is trained using the surrogate gradient method. 

We propose a novel approach for single-channel SE in real-world noise conditions using a supervised SNN framework and a U-Net inspired architecture with direct input mapping. The current work presents several important contributions toward the development of SNNs for SE that address limitations of prior studies.

Firstly, we propose an SNN-based model for SE, which learns to map the logarithmic power spectrum (LPS) of noisy speech to that of clean speech. This represents a significant advancement in the field, as to the best of our knowledge, this is the first attempt to apply an SNN architecture to the SE task using a direct mapping strategy instead of masking. Secondly, we employ a direct encoding approach that allows the SNN to simultaneously encode acoustic features of speech and suppress noise, which has not been explored in previous SNN-based SE models. Thirdly, we use trainable neuron parameters (decay strength and membrane threshold) for enhanced learning. Finally, the performance of the proposed SNN model is compared to an ANN model with a similar architecture, demonstrating slightly lower but still comparable performance.

This work further highlights the potential of SNNs in the SE task and supports the notion that SNNs can provide a viable energy-efficient alternative to conventional ANN-based models. Overall, it contributes towards advancing the development of SNNs for SE applications, with potential implications for the broader field of speech processing and communication technologies.

\section{Background}\label{sec:background}

\subsection{Spiking neural network}\label{ssec:snn}

SNNs are a type of neural network that model the behavior of biological neurons \cite{Long}. Unlike traditional ANNs that use continuous valued activation functions, SNNs use discrete spikes to communicate information between neurons. The dynamics of SNNs are governed by the spiking behavior of individual neurons, which emit a spike when their membrane potential reaches a threshold. The spike trains then propagate to post-synaptic neurons, where they contribute to the computation of post-synaptic currents.

The Leaky Integrate-and-Fire (LIF) neuron model with current-based synapses \cite{Gerstner} is a mathematical model widely used in SNNs due to its simplicity and computational efficiency. The dynamics of a LIF neuron $i$ in the layer $l$ can be described as:
\begin{equation}
    \begin{aligned}
    I_i^{(l)}(t+1) = \alpha I_i^{(l)}(t) + \sum _j W_{ij}^{(l)} S_{j}^{(l-1)}(t) + \sum _j V_{ij}^{(l)} S_{j}^{(l)}(t)
    \end{aligned}
    \label{eqn:in_curr}
\end{equation}
\begin{equation}
    U_i^{(l)}(t+1) = \beta U_i^{(l)}(t) + I_i^{(l)}(t) - U_{th} S_i^{(l)}(t)
    \label{eqn:mem_pot}
\end{equation}
\begin{equation}
    S_i^{(l)}(t) = \Theta (U_i^{(l)}(t) - U_{th})
    \label{eqn:spike}
\end{equation}
where $I_i^{(l)}(t)$ represents the input current at $t$, $U_i^{(l)}(t)$ represents the membrane potential at $t$, $S_i^{(l)}(t)$ represents the spike train at $t$, $W_{ij}^{(l)}$ and $V_{ij}^{(l)}$ are synaptic weight matrices for the feedforward and recurrent connections, respectively. $\alpha = exp(- \frac{\Delta_t}{\tau_{syn}})$ and $\beta = exp(- \frac{\Delta_t}{\tau_{mem}})$ are the decay strengths for the input current and the membrane potential, respectively.

The spike behavior of individual neurons is described by Equation \ref{eqn:spike}, where $\Theta$ is the Heaviside step function that outputs a spike when the membrane potential of the neuron exceeds a threshold $U_{th}$.

\subsection{Surrogate Gradient}

Surrogate gradients have recently emerged as a powerful tool for training SNNs directly using backpropagation \cite{Neftci, Zenke}. In traditional ANNs, backpropagation computes the gradient of the loss function with respect to the weights of the network using the chain rule. However, this method cannot be directly applied to SNNs because the spiking behavior of neurons is non-differentiable, making it impossible to compute gradients through the spiking activation function. Surrogate gradients provide an alternative way to compute gradients in SNNs by approximating the non-differentiable spiking activation function with a differentiable function. This allows for the use of backpropagation to update the weights of the network during training.

\section{Method}\label{sec:method}

\subsection{System overview}

\begin{figure}[htbp]
\centerline{\includegraphics[scale=0.35]{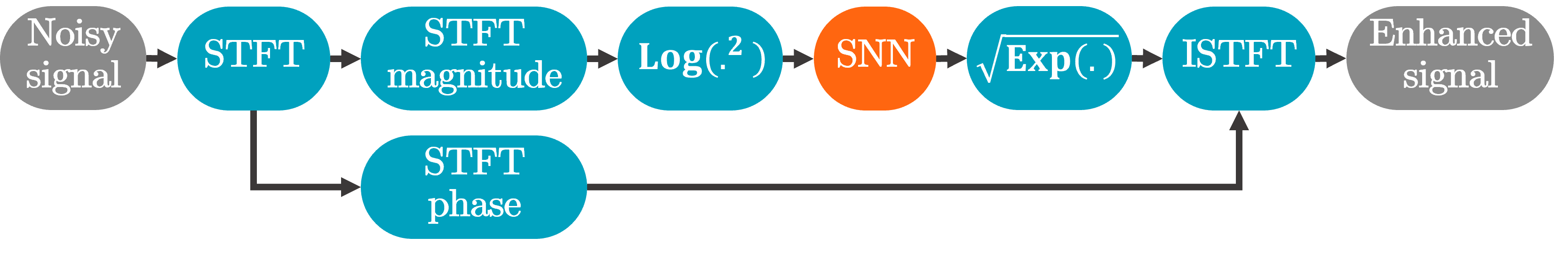}}
\caption{Speech enhancement system overview.}
\label{fig:system_overview}
\end{figure}

In this work, we propose an SNN-based approach for SE as depicted in Figure \ref{fig:system_overview}. The proposed system operates in the time-frequency domain, where the STFT is used to decompose the input speech signal into its constituent spectral components. Subsequently, the magnitude of the STFT is computed and transformed into LPS using the logarithmic and power functions. The SNN model is trained to learn a non-linear mapping from the noisy input LPS to an enhanced LPS, leveraging the inherent non-linear properties of the underlying speech signal. The inverse transform of the SNN output is computed using the exponential and square root operations, yielding the enhanced STFT magnitude. Finally, the enhanced STFT magnitude is combined with the noisy STFT phase to reconstruct the enhanced speech signal.

\subsection{Direct input encoding}\label{ssec:encoding}

In this work, we adopt the direct encoding approach \cite{Rueckauer, Lu, Rathi}. We apply the noisy LPS directly as input to the SNN, which is then directly encoded into binary spikes using the LIF neurons within the first layer of the network. The direct encoding method has been shown to improve activation sparsity and energy efficiency of SNNs, as reported in \cite{Rathi}. It provides better control of input information flow by attenuating irrelevant inputs and increasing activation sparsity.

\subsection{Model architecture}

\begin{figure*}[htbp]
\centerline{\includegraphics[scale=0.7]{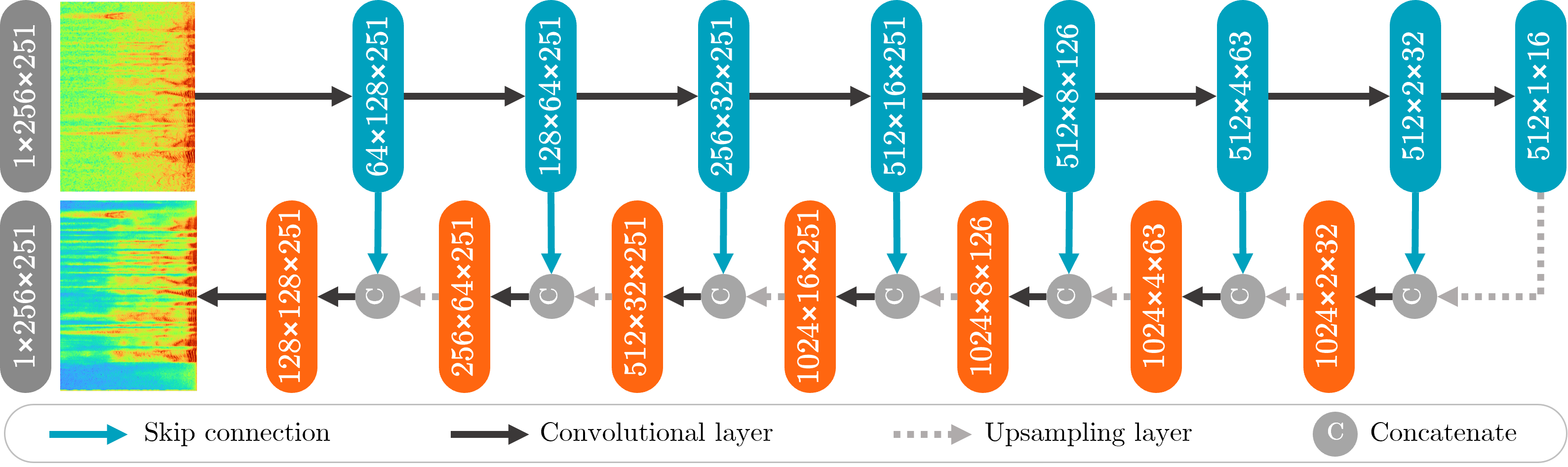}}
\caption{Proposed SNN architecture.}
\label{fig:snn_architecture}
\end{figure*}

Recently, the U-Net architecture, which is widely employed in image segmentation \cite{Ronneberger}, has been adapted for the task of SE using ANNs \cite{Pascual, Macartney, Bulut}. The proposed SNN architecture (shown in Figure \ref{fig:snn_architecture}) is composed of layers arranged in a U-shaped configuration, with skip connections between the encoder and decoder. The encoder component of the network extracts relevant features from the input LPS, whereas the decoder generates the enhanced LPS.

The SNN takes as input the LPS and consists of eight encoder layers, followed by seven decoder layers, and a final readout layer. Downsampling within the encoder is accomplished via strided convolutions. The decoder employs the neuromorphic-friendly nearest neighbor method for upsampling. The readout layer is composed of non-spiking neurons. 

The neurons in this SNN are modeled using LIF neurons, which emulates the fundamental behavior of biological neurons. The LIF neurons accumulate input signals over time and emit a spike when the membrane potential reaches a predefined threshold. To enable efficient training of our SNN using backpropagation, we use a differentiable approximation of the spiking activation function, which is the Heaviside step function. In this work, we use the derivative of the ArcTan function as a surrogate function. This approach enables us to efficiently optimize the parameters of our SNN using gradient-based methods while preserving the spiking behavior of the LIF neurons.

\subsection{Loss function}

We use the log-spectral distance (LSD) loss:
\begin{equation}
    \begin{aligned}
    L_{LSD} = \frac{1}{M} \sum_{m=0}^{M-1} \sqrt{\frac{1}{K} \sum_{k=0}^{K-1} (X[m, k] - \hat{X}[m, k])^2}
    \end{aligned}
    \label{eqn:in_curr}
\end{equation}
where $X[m, k]$ and $\hat{X}[m, k]$ represent, respectively, the clean and estimated LPS for the $m^{th}$ time frame and $k^{th}$ frequency bin.

\section{Experiments}\label{sec:experiments}

\subsection{Dataset}

In this study, we utilized a publicly available dataset \cite{Valentini} \cite{Valentini_link}, which comprises recordings of both clean speech and noisy speech, with various types of noise added at different SNRs. The speech data is characterized by a gender-balanced composition of English speakers, with the clean speech recordings of sentences extracted from the voice bank corpus (VCTK) \cite{Veaux} with a sampling frequency of 48 kHz.

The training set of the dataset consists of approximately $10$ hours of speech signals from $28$ speakers, with $10$ different types of noise added at SNR values of $15$, $10$, $5$, and $0$ dB. The types of noise include two artificial and eight real-world noise recordings obtained from the Demand dataset \cite{Thiemann}. On the other hand, the test set comprises $30$ minutes of speech signals from two speakers, with five types of real-world noise from the Demand dataset added at SNR values of $17.5$, $12.5$, $7.5$, and $2.5$ dB. We randomly split the training set into training and validation sets.

\subsection{Data preprocessing}

In this study, a standard preprocessing procedure was applied, which included downsampling of input speech signals in the temporal domain to $16$ kHz. STFT was computed on the speech signals using a frame length of $32$ ms with a hop length of $16$ ms and Hann window.

\subsection{Training setup}

To train the model, we use the Adam optimizer \cite{Kingma} with a learning rate of $0.002$, decay rates $\beta_1 = 0.5$ and $\beta_2 = 0.9$, and a batch size of $32$ for $60$ epochs. Weights of the convolutional layers are initialized from a normal distribution with zero mean and $0.2$ standard deviation. Decay strengths and membrane threshold of the LIF neurons are initialized from a normal distribution with mean values of $0.05$ and $1.0$, respectively, and both with a standard deviation of $0.01$.

\subsection{Evaluation metrics}

Evaluating the performance of a SE system is a critical aspect of determining its efficacy in real-world scenarios. In this work, we propose the use of three widely used objective metrics to evaluate the performance of the proposed SNN-based SE system.

The first metric used is the Perceptual Estimation of Speech Quality (PESQ) \cite{Rix}, which has been widely adopted in SE literature to evaluate the similarity between the enhanced speech signal and the original speech signal. PESQ is a perceptual metric that is intended to correlate well with human perception of speech quality. It ranges from $-0.5$ to $4.5$. The second metric used is the Short Term Objective Intelligibility (STOI) \cite{Taal}. STOI is a measure of the intelligibility of the speech signal and has been shown to be highly correlated with human speech intelligibility judgments. It ranges from $0$ to $1$. Lastly, we used the Deep Noise Suppression Mean Opinion Score (DNSMOS) ANN-based estimator \cite{Reddy}. DNSMOS comprises three scores, namely, the quality of the speech signal (SIG), the background noise (BAK), and the overall quality of the enhanced speech (OVRL). These scores range from $1$ to $5$ For all the metrics used, a higher score indicates a better performance of the algorithm.

\section{Results}\label{sec:results}

In this section, we present the results of the proposed SNN-based speech enhancement approach and compare it against classical and state-of-the-art approaches. Table \ref{tab:results} summarizes the numerical results obtained from our experiments using the aforementioned metrics.

\begin{table}[htbp]
\caption{Evaluation metrics comparison}
\label{tab:results}
\begin{center}
    \begin{tabular}{|c|c|c|c|c|c|}
    \hline
    \textbf{} & \textbf{} & \textbf{} & \multicolumn{3}{|c|}{\textbf{DNSMOS}} \\
    \cline{4-6} 
    \textbf{System} & \textbf{PESQ} & \textbf{STOI} & \textbf{\textit{OVRL}} & \textbf{\textit{SIG}} & \textbf{\textit{BAK}} \\
    \hline
    Noisy                           & 1.97  & 0.92  & 2.69 & 3.34 & 3.12 \\
    Wiener      \cite{Lim}          & 2.22  & - & - & - & - \\
    SEGAN       \cite{Pascual}      & 2.16  & - & - & - & - \\
    Wave-U-Net  \cite{Macartney}    & 2.40  & - & - & - & - \\
    MMSE-GAN    \cite{Williamson}   & 2.53  & - & - & - & - \\
    D+M         \cite{Yao}          & 2.73  & - & - & - & - \\
    U-Net       \cite{Bulut}        & 2.90  & 0.93  & - & - & - \\
    Equivalent ANN                  & 2.89  & 0.94  & 2.92 & 3.23 & 3.90 \\
     \hline
    SDNN        \cite{Timchek}      & 2.00  & 0.91 & 2.44 & 3.05 & 3.09 \\
    Proposed                        & 2.66  & 0.92 & 2.81 & 3.13 & 3.85 \\
    \hline
    \multicolumn{5}{l}{}
    \end{tabular}
\label{tab1}
\end{center}
\end{table}

Primarily, the proposed approach demonstrates significant improvements across most of the evaluation metrics when compared to the unprocessed noisy speech signals, indicating its effectiveness in SE. Furthermore, the outcomes of the performance evaluation demonstrate that the proposed SNN approach surpasses several state-of-the-art techniques in terms of PESQ and has a comparable STOI value.

Moreover, we present an evaluation of the proposed SNN model in comparison with a recently published SNN-based approach, specifically the baseline SDNN model introduced in \cite{Timchek} for the Intel N-DNS Challenge. To ensure benchmarking consistency, we train and evaluate the SDNN model using the same dataset as the proposed model. The proposed U-Net based SNN significantly outperforms the SDNN baseline across all evaluation metrics. 

It is worth noting that the authors of \cite{Timchek} report much better results in their original paper, however, the SDNN model was trained in \cite{Timchek} on a large corpus of $500$ hours of audio data, whereas the dataset utilized in our study comprises only around $10$ hours of audio. Therefore, our proposed approach seems to overcome the limitations of dataset size and variability. These findings attest to the efficacy of the proposed method for SE and suggest its potential for real-world applications.

Additionally, we compare the proposed SNN architecture with an ANN having an equivalent architecture. Our findings indicate that the proposed SNN approach achieves slightly lower but still comparable performance in terms of DNSMOS, demonstrating its efficacy in enhancing speech signals while employing fewer computational resources than the ANN-based approach.

It is important to note that while the proposed SNN approach has exhibited competitive performance for speech enhancement, it is essential to conduct further investigations to explore its performance under diverse frameworks, such as different speech encoder, neuron model, loss function, or using a masking based method. Such investigations would help to elucidate the effectiveness of the proposed approach across a broader range of scenarios and enable a more thorough understanding of its capabilities and limitations.

\section{Conclusion}\label{sec:conclusion}

In this paper, we proposed a novel single-channel speech enhancement approach based on a U-Net SNN architecture. We show that SNN can perform large scale regression tasks such as speech enhancement. The objective evaluation results show that the proposed approach outperforms many state-of-the-art ANN-based methods. Also, it outperforms the Intel N-DNS Challenge SDNN baseline solution. Furthermore, we achieved competitive results in comparison with an equivalent ANN architecture, indicating the promising potential of SNNs for speech enhancement applications. Overall, the present study highlights the potential of SNNs in the domain of speech enhancement and provides a valuable contribution to the ongoing research efforts in this area.

\balance

\end{document}